%% file: fpcp3.tex
\documentclass[12pt]{article}
 \usepackage{epsfig}
\input econfmacros.tex
 \def\beq{\begin{equation}}
 \def\eeq{\end{equation}}
 \def\bea{\begin{eqnarray}}
 \def\eea{\end{eqnarray}}
 \def\nn{\nonumber}
 \def\sss{\scriptscriptstyle}
 
 \def\bd{B_d^0}
 \def\bdbar{{\bar B}_d^0}

 \def\barp{{\raise.35ex\hbox
 {${\sss (}$}}---{\raise.35ex\hbox{${\sss )}$}}}
 \def\bdbarp{\hbox{$B_d$\kern-1.4em\raise1.4ex\hbox{\barp}}}
 \def\bsbarp{\hbox{$B_s$\kern-1.4em\raise1.4ex\hbox{\barp}}}

 \def\barpk{{\raise.35ex\hbox
 {${\sss (}$}}--{\raise.35ex\hbox{${\sss )}$}}}
 \def\kbarp{\hbox{$K$\kern-0.9em\raise1.4ex\hbox{\barpk}}}
 \def\roughly#1{\mathrel{\raise.3ex\hbox
 {$#1$\kern-.75em\lower1ex\hbox{$\sim$}}}}

 \def\adirpm{{a_{\sss dir}^{+-}}}
 \def\adir00{{a_{\sss dir}^{00}}}
 
 \def\alphaeff{{\alpha_{\sss eff}}}

 \def\B00{B^{00}}
 \def\Bp0{B^{+0}}

\def\Title#1{\begin{center} {\Large {\bf #1} } \end{center}}

\begin{document}

\begin{flushright}
IMSc-2002/06/13\\
\end{flushright}

\Title{Penguin Pollution in $\bd \to \pi\pi$\footnote{Talk given at {\it Flavor Physics and CP Violation (FPCP)}, Philadelphia,
PA, USA, May 2002, major part of this talk is based on work done in collaboration with Michael Gronau, David London and Rahul Sinha~\cite{GLSS}.}
}

\bigskip\bigskip


\begin{raggedright}  

{\it Nita Sinha\index{Sinha, N.}\\
The Institute of Mathematical Sciences\\
Taramani, Chennai 600113, INDIA}
\bigskip\bigskip
\end{raggedright}

A principal decay mode considered for the measurement of the angle
$\alpha$, is $\bd(t) \to \pi^+ \pi^-$. Unfortunately, this mode
suffers from a well-known problem: penguin contributions may be
large~\cite{penguins}, and their presence will spoil the clean
extraction of $\alpha$. In the presence of penguin amplitudes, the CP
asymmetry in $\bd(t)\to\pi^+\pi^-$ does not measure $\sin 2\alpha$,
but rather some effective (``polluted'') angle $2\alphaeff$.  We may write
the time-dependent rate of $\bd(t)\to\pi^+\pi^-$ as,
\bea
\Gamma(B^0(t)\to\pi^+\pi^-) = e^{-\Gamma t}B^{+-}\left [1 + \adirpm
\cos(\Delta mt) - y\sin 2\alphaeff\sin(\Delta mt)\right ]~,\nn\\
{\rm where},~~
B^{+-}\equiv{1\over 2} \left( |A^{+-}|^2+|{\bar A^{+-}}|^2 \right) ~, ~~~~~~~\adirpm\equiv{{|A^{+-}|^2-|{\bar A^{+-}}|^2} \over
{|A^{+-}|^2+|{\bar A^{+-}}|^2} }~,
\label{observables}
\eea
$A^{+-}$ and ${\bar A^{+-}}$ are the amplitudes for $\bd \to\pi^+\pi^-$
 and $\bdbar \to \pi^+\pi^-$, respectively, and $y \equiv\sqrt{1-(\adirpm)^2}$.
Writing the time dependent CP asymmetry as,
\begin{equation}
{\cal A} = C_{\pi\pi}\cos(\Delta mt) + S_{\pi\pi}\sin(\Delta mt) ~,
\label{asym}
\end{equation}
we have, $C_{\pi\pi}=\adirpm$ and $S_{\pi\pi}=-y\sin 2\alphaeff$.
In Eq.~(\ref{observables}), the coefficient of the $\sin(\Delta M t)$ term,
probes the relative phase between the $A^{+-}$ and $e^{-2i\beta}{\bar
  A}^{+-}$ amplitudes, and this phase, $2\alphaeff=2\alpha$, in
absence of penguin contributions.

The problem of penguin pollution can be eliminated with
the help of an isospin analysis~\cite{isospin}. By measuring the rates
for $B^+ \to \pi^+ \pi^0$ and $\bd/\bdbar \to \pi^0\pi^0$, in addition
to $\bd(t)\to\pi^+\pi^-$, $\alpha$ can again be measured cleanly.

However, the isospin analysis requires separate measurement of $BR(\bd
\to \pi^0\pi^0)$ and $BR(\bdbar \to \pi^0\pi^0)$, and therefore suffers
from potential practical complications: (i)The branching ratio for
$\bd\to \pi^0\pi^0$ is expected to be smaller than $\bd\to\pi^+\pi^-$.
(ii)The presence of two $\pi^0$'s in the final state means that the
reconstruction efficiency is smaller. (iii)It will be necessary
to tag the decaying $\bd$ or $\bdbar$ meson, which further reduces the
measurement efficiency.  Hence, we may only have, an actual measurement
or an upper limit, on the sum of the branching ratios. In this case, a
full isospin analysis cannot be carried out.

Question: assuming that we have, at best, only partial knowledge of
the sum, ($BR(\bd \to \pi^0\pi^0)+BR(\bdbar \to \pi^0\pi^0)$), can we
at least put bounds on the size of penguin pollution? In the presence
of penguin amplitudes, the CP asymmetry in $\bd(t)\to\pi^+\pi^-$
measures $\sin 2\alphaeff$. Writing $2\alphaeff=2\alpha + 2\theta$,
where $2\theta$ parametrizes the effect of the penguin contributions,
the more precise question: is it possible to constrain $\theta$?  As
demonstrated by Grossman and Quinn(GQ)~\cite{GQ} and later by
Charles~\cite{Charles}, the answer to this question is yes.  They were
able to show that $|2\theta|$ can be bounded even if we have only an
upper limit on the sum of $BR(\bd \to \pi^0\pi^0)$ and $BR(\bdbar \to
\pi^0\pi^0)$:
\begin{equation}
\cos 2\theta \ge {1 - 2 B^{00} / B^{+0} \over y} ~,~~~~\cos 2\theta \ge {1 - 4 B^{00} / B^{+-} \over y} ~,
\label{GQbound}
\end{equation}
where, $B^{00}$ and $B^{+0}$ are defined analogous to the definition
of $B^{+-}$ in Eq.~(\ref{observables}).  Next question: does a more
stringent bound exist? The answer to this is also yes; the most
stringent bound possible on $|2\theta|$ was obtained by Gronau,
London, Sinha and Sinha~\cite{GLSS}, by requiring that the two isospin
triangles close and have a common base.  We note, however, that
neither of the bounds in Eq.~(\ref{GQbound}) involves all three
charge-averaged decay rates, $B^{+-}$, $B^{+0}$ and $B^{00}$.  Thus, a
condition for the closure of the two isospin triangles is not included
in these bounds.

\begin{figure}[htb]
\begin{center}
\epsfig{file=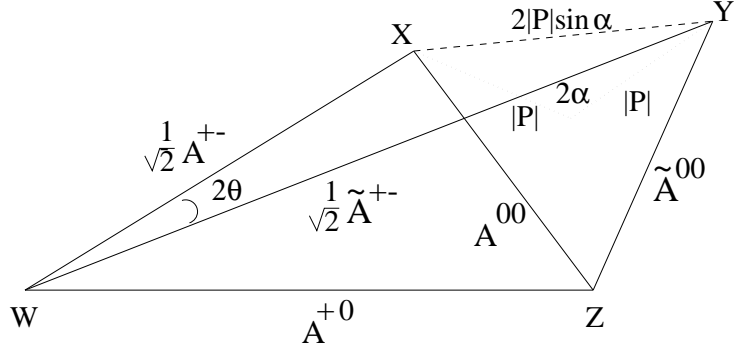,height=1.8in}
\caption{The $A$ and ${\tilde A}$ isospin triangles.}
\label{isotriangles}
\end{center}
\end{figure}

We now present a geometrical derivation of this new bound on
$|2\theta|$.  We assume that the charge-averaged rates $B^{+-}$ and
$B^{+0}$ have been measured, and that we have (at least) an upper
bound on $B^{00}$. %
The $B\to\pi\pi$ decay amplitudes take the form
\begin{equation}
{1 \over \sqrt{2}} A^{+-}  =  T e^{i\gamma} + P e^{-i\beta} ~,~~~~A^{00} = C e^{i\gamma} - P e^{-i\beta}~,~~~~A^{+0} = (C + T) e^{i\gamma},
\label{amps}
\end{equation}
where, the complex amplitudes $T$, $C$ and $P$, which are sometimes
referred to as ``tree", ``colour-suppressed" and ``penguin"
amplitudes, include strong phases.  Note that we have implicitly
imposed the isospin triangle relation,
\begin{equation}
{1 \over \sqrt{2}} A^{+-} + A^{00} = A^{+0} ~.
\end{equation}
The ${\bar A}$ amplitudes can be obtained from the $A$ amplitudes by
reversing the signs of the weak phases.
It is convenient to define the new amplitudes ${\tilde A}^{ij} \equiv
e^{2 i \gamma} {\bar A}^{ij}$. Then, ${\tilde A}^{-0} = A^{+0}$, so that the $A$ and ${\tilde A}$
triangles have a common base. (A tiny electroweak penguin amplitude,
forming a very small angle between $A^{+0}$ and ${\tilde A}^{-0}$, will
be neglected here.) In the absence of penguin contributions,
${\tilde A}^{+-} = A^{+-}$, thus, the relative phase $2\theta$ between
these two amplitudes is due to penguin pollution.  Also, the relative
phase between the penguin contributions in ${\tilde A}^{00}$ and
$A^{00}$ is $2(\beta + \gamma) \sim 2\alpha$. This information is
encoded in Fig.~\ref{isotriangles}. Note that the distance between the
points $X$ and $Y$ is $2 \ell \equiv 2 |P| \sin\alpha$.
Now, $|P|$ can be expressed in terms of observables~\cite{Charles}, and we can therefore write,
\begin{equation}
\ell = {1 \over 2} \sqrt{B^{+-}} \sqrt{1 - y \cos 2\theta} ~.
\label{elldef}
\end{equation}
Thus, a constraint on $\ell$ implies a bound on $\cos 2\theta$.

In order to constrain $\ell$, we proceed as follows. First, we assign
a coordinate system to Fig.~\ref{isotriangles} such that the origin is
at the midpoint of the points $X$ and $Y$. The points $X$, $Y$, $W$ and $Z$
correspond respectively to the coordinates $(+\ell, 0)$, $(-\ell,
0)$, $(x_1, y_1)$
and $(x_2, y_2)$. The goal of the exercise is to find the values of
the coordinates $(x_1, y_1)$
and $(x_2, y_2)$. We then note that
\begin{equation}
\begin{array}{ll}
B^{+-}=2 (x_1^2 + y_1^2) + 2 \ell^2,&B^{+-} a_{dir}^{+-} = - 4 x_1 \ell ~,\nn\\
B^{00}= (x_2^2 + y_2^2) + \ell^2,&B^{+0}= (x_1^2 + y_1^2) + (x_2^2 + y_2^2) - 2 x_1 x_2 - 2 y_1 y_2 ~.
\end{array}
\end{equation} 
We therefore have four (nonlinear) equations in four unknowns, and we
can solve for these coordinates as a function of $\ell$.
However, we must obtain {\it only real solutions} for $x_2$ and $y_2$,
otherwise the triangles do not close.  This puts a constraint on
$\ell$, which in turn, gives the following bound,
\begin{equation}
\cos 2\theta \ge { \left( {1\over 2}B^{+-} + B^{+0} - B^{00} \right)^2 -
  B^{+-} B^{+0} \over B^{+-} B^{+0} y} ~.
\label{superbound}
\end{equation}
This is the new lower bound on $\cos 2\theta$ (or upper bound on
$|2\theta|$).

The new bound contains the two previous bounds as limiting cases.  We
can rewrite this lower bound on $\cos 2\theta$ in two alternate
forms~\cite{GLSS}, which involve a sum of two terms. In one form, the
first term is simply the GQ bound and in the other, it is the Charles
bound. The second terms in both forms are positive definite. Hence,
the new bound is stronger than the GQ as well as the Charles bound,
and since, all isospin information has been used in obtaining
Eq.~(\ref{superbound}), this is {\it the most stringent possible bound
  on $\cos 2\theta$}.

One would also like to know if it is possible to find a lower
bound on $|2\theta|$?  Unfortunately, the answer is no. This can be
seen quite clearly in Fig.~\ref{isotriangles}. Suppose that the
two-triangle isospin construction can be made for some nonzero value
of $2\theta$.  It is then straightforward to show that one can always
rotate $A^{+-}$ and ${\tilde A}^{+-}$ continuously around $W$ towards
one another, without changing $B^{00}$, until they lie on one line
corresponding to $\theta = 0$.  Thus, without measuring separately
$\bd\to\pi^0\pi^0$ and $\bdbar\to\pi^0\pi^0$, one cannot put a lower
bound on the penguin pollution parameter.

Using the world average values~\cite{waFPCP}, $BR(\bd \to
\pi^+\pi^-)=5.2\pm0.6$ and $BR(\bd \to \pi^+\pi^0)=4.9\pm1.1$ and
Babar's value of $BR(\bd \to \pi^0\pi^0)=0.9^{+0.9+0.8}_{-0.7-0.6}$,
our bound yields, $\theta<57^\circ$ or $\theta>123^\circ$ at $90\%$ CL, while the GQ
bound gives $\theta< 61^\circ$ or $\theta>119^\circ$ at $90\%$ CL\footnote{We thank Andreas Hoecker and Rainer Bartoldus for help in estimating
the numerical values of our bound.}. The Charles bound gives weaker
constraints. Note that these values are obtained using zero direct
asymmetry, the bounds will be stronger if direct asymmetry is
non-vanishing.

The bound on $\cos 2\theta$ in Eq.~(\ref{superbound}) together with the condition that $\cos 2\theta \le 1$, leads to a lower limit on
$B^{00}/B^{+-}$. This lower limit, as well as an upper limit on
the same quantity, follows directly from the closure of the two
isospin triangles, which can be shown to imply that
\begin{equation}
{1\over 2} + {B^{+0} \over B^{+-}} - \sqrt{ {B^{+0} \over B^{+-}} (1 +
  y) } \le {B^{00}\over B^{+-}} \le {1\over 2} + {B^{+0} \over B^{+-}}
+ \sqrt{ {B^{+0} \over B^{+-}} (1 + y) } ~.
\label{B00bounds}
\end{equation}
The limits are weakest for $y=1$. Using the central values of the world averages
listed above, one finds $0.069 \le B^{00}/B^{+-} \le 2.815$,
for $\adirpm=0$; again, a non-zero value of the direct asymmetry will raise
the lower limit.
This lower limit on $B^{00}/B^{+-}$ is useful, as 
it will give experimentalists some knowledge of the branching ratios
for $\bd/\bdbar \to \pi^0\pi^0$, and thus will help to anticipate
the feasibility of the full isospin analysis. In addition, since the bound
on $B^{00}/B^{+-}$ relies only on the closure of the two triangles, it
will hold even in the presence of isospin-violating
electroweak-penguin contributions. However, it has been
pointed out by Gardner~\cite{Gardner} that the triangles will not
close in the presence of other isospin-violating effects such as
$\pi^0$--$\eta,\eta'$ mixing. A comparison of the actual
branching ratio $B^{00}$ with this bound, may therefore give some information
about the size of such isospin-violating effects.

Although no lower limit can be obtained on the penguin-pollution angle
$|2\theta|$, we note that a lower bound can be derived for the
magnitude of the penguin amplitude $P$ from measurements of
$\bd(t)\to\pi^+\pi^-$ alone,
\begin{equation}
|P|^2_{min} = {B^{+-} (1 - y^2) \over 4 (1 - y \cos 2\alphaeff)} ~.
\end{equation}

Recently Belle~\cite{Belle} and Babar~\cite{Babar} announced their
results for the CP violating asymmetries $C_{\pi\pi}$ and
$S_{\pi\pi}$. These asymmetries, may be written~\cite{GR} in terms of
the three parameters $|P/T|$, the strong phase difference of the
penguin and tree amplitudes, $\delta=\delta_P-\delta_T$ and the weak
phase $\alpha$. If one assumes a value of $|P/T|$, one can
determine $\alpha$ upto discrete ambiguities from the time dependent
study of the $\bd(t) \to \pi^+ \pi^-$ mode alone.  There are discrete
ambiguities associated with mapping the observables $(S_{\pi\pi},
C_{\pi\pi})$ with the parameters $(\alpha,\delta)$. The ambiguities
can be resolved by a measurement of $R_{\pi\pi}$, which is the ratio
of the flavor-averaged $\bd \to \pi^+ \pi^-$ branching ratio to its
predicted value due to the tree amplitude alone.
If one averages over Belle and Babar asymmetry results, then a larger
value of $\alpha$ is favored~\cite{GR}.

%
Concluding remark: By the end of this summer, with improved statistics
($\approx 100fb^{-1}$), we hope to have a much clearer understanding about the
amount of penguin pollution in $\bd\to\pi^+\pi^-$.

\bigskip
N.S. would like to thank the organizers of FPCP for a great
conference. She thanks David London for the hospitality of the
Universit\'e de Montr\'eal, where some of this work was done. This
work is supported by the young scientist award from the Department of
Science and Technology, India.

\end{document}

%% file: econfmacros.tex
\def\beq{\begin{equation}}
\def\eeq#1{\label{#1}\end{equation}}
\def\eeqn{\end{equation}}

\def\beqa{\begin{eqnarray}}
\def\eeqa#1{\label{#1}\end{eqnarray}}
\def\eeqan{\end{eqnarray}}

\let\bar=\overbar

\def\Dslash{\not{\hbox{\kern-4pt $D$}}}
\def\dslash{\not{\hbox{\kern-2pt $\del$}}}

\def\msb{{\bar{\ssstyle M \kern -1pt S}}}

